\documentclass[twocolumn]{autart}    
\usepackage{hyperref}
\usepackage{comment}

\usepackage[utf8]{inputenc}
\usepackage[T1]{fontenc}
\usepackage{graphicx}
\usepackage{epsfig}
\usepackage{subfigure}
\usepackage{amsmath}
\usepackage{amsfonts}
\usepackage{amssymb} 
\usepackage{epstopdf}
\usepackage{subfigure}
\usepackage{xcolor}
\begin{document}

\begin{frontmatter}

\title{Perturbed Integrators Chain Control via Barrier
Function Adaptation and Lyapunov Redesign}
\thanks[footnoteinfo]{Corresponding author}
\author[a]{Manuel A. Estrada}
\author[b]{Claudia~A.~P\'erez-Pinacho}
\author[b]{Christopher~D.~Cruz-Ancona\thanksref{footnoteinfo}}
\author[a]{Leonid~Fridman}


\address[a]{Facultad de Ingenier\'ia, Universidad Nacional Aut\'onoma de M\'exico (UNAM), Ciudad de M\'exico 04510, M\'exico (manuelestrada@comunidad.unam.mx; lfridman@unam.mx)} 
\address[b]{ Tecnologico de Monterrey, School of Engineering and Sciences, Ave Eugenio Garza Sada 2501, Monterrey, Nuevo Leon, 64849 (christopher.cruz.ancona@tec.mx; caperezpinacho@tec.mx) } 


\begin{abstract}
{Lyapunov redesign is a classical technique that uses a nominal control and its corresponding nominal Lyapunov function to design a discontinuous control, such that it compensates the uncertainties and disturbances}. In this paper, the idea of Lyapunov redesign is used to propose an adaptive time-varying gain controller to stabilize a class of perturbed chain of integrators with an unknown control coefficient. It is assumed that the upper bound of the perturbation exists but is unknown.  A proportional navigation feedback type gain is used to drive the system's trajectories into a prescribed vicinity of the origin in a predefined time,  measured using a quadratic Lyapunov function.  Once this neighborhood is reached, a barrier function-based gain is used, ensuring that the system's trajectories never leave this neighborhood despite uncertainties and perturbations.  Experimental validation of the proposed controller in Furuta's pendulum is presented.
\end{abstract}

\end{frontmatter}
\section{Introduction}
{Lyapunov redesign is a classical robustification technique in which an additional discontinuous control is designed to compensate matched uncertainties and external disturbances. Such approach uses the knowledge of a nominal control and its corresponding Lyapunov function to design an appropriate switching manifold. The discontinuous control using such manifold ensures the negativity of the nominal Lyapunov function if the upper-bound of the perturbations is known \cite{Leitmann1979}\cite{Gutman1979}}. Since the upper bound of the perturbation is usually unknown or overestimated, adaptive gains in controllers are needed to stabilize the system \cite{corless84adap}\cite{shtessel2016adaptive}. In adaptive strategies for systems with an unknown upper bound of the perturbation and unknown control gain, three different issues should be solved simultaneously:

\begin{enumerate}
\item[(i)] To confine the trajectories of the system into a prescribed neighborhood of the origin before a predefined time moment.
\item[(ii)] To update the controller's gain, once a prescribed neighborhood is reached, the system's trajectories are confined in this region while the controller signal compensates for the perturbation.
\item[(iii)] To produce a bounded continuous control signal.
\end{enumerate}

Recently, explicit time-varying controllers have been proposed in \cite{song2017,gomez2020,aldana2021,chitour2020} that can be used to solve the problem (i) for the stabilization of a perturbed chain of integrators. The first approach of this type \cite{song2017}, so-called Proportional Navigation Feedback (PNF), ensures convergence in prescribed time with a proportional controller with a time-varying gain that becomes  unbounded as the solution tends to zero at the prescribed time, that is why an absolutely continuous solution might not be defined at the convergence time, but in a generalized sense  \cite{seeber2023generalized}. Consequently, adding measurement noise worsen regulation accuracy at the prescribed time, see  \cite{aldana2022,chitour2020}. 

To solve the problem (i), the works \cite{gomez2020,aldana2021,chitour2020} take advantage of the time-varying gains to redesign the fixed-time controllers ensuring the convergence to the origin before the predefined convergence time 
moment and maintain the solutions therein for all future times. However, the usage of these controllers does not consider the case when the control coefficient is uncertain, and to maintain the solution in the origin, these controllers require the knowledge of the upper bound of perturbations. 

Another strategy is to drive the system's trajectories to an arbitrary neighborhood of the origin before the predefined time with time-varying feedback \cite{cao2021} and stay therein for all future times.  The paper \cite{cao2021} uses a perturbation estimator in combination with a performance function-based controller to render a {relative degree two} system  practical stabilizable within an assigned reaching time.  The paper \cite{cao2021} considers parametric uncertainty; its effectiveness relies on an Artificial Neural Network, which can be computationally expensive, and this approach is developed only for systems of order two. Thus, ensuring the problem (i) solution for \textit{arbitrary relative degree} perturbed systems is still an open problem. 

There are a lot of sliding mode approaches \cite{plestan10,plestan13,shtessel2016adaptive,oliveira16,edwards2016adaptive} to solve the problem (ii)  keeping the trajectories in a neighborhood of the sliding set, and then fixing \cite{negrete16,Incremona2016} or reducing \cite{plestan13,shtessel12} the gains of the controller. However, in such approaches, a control law is discontinuous, which produces the \textit{chattering} effect, high-frequency oscillations that may damage systems.
Other approaches ensure a prescribed neighborhood of a sliding set via barrier functions \cite{obeid18,Obeid2020}, or monitoring functions \cite{Hsu2019,Oliveira2017}. 
An important drawback of the previous works is that, due to the absence of knowledge on the upper bound of the perturbations, the reaching phase for arrival into a {neighborhood of the origin of the state space} cannot be predefined. 

In the paper \cite{cruz2021barrier}, it is shown that a barrier function-based adaptation of Lyapunov redesign is not straightforward because to ensure the property (i) it is necessary to show that the system’s trajectories converge uniformly into a priori predeﬁned vicinity of origin for all system’s states in a predefined time. Whether convergence to a neighborhood of a sliding set can be ensured by using a uniform reaching phase strategy in \cite{cruz2023uniform}, the uniform convergence from the point of arrival on the sliding set to the predefined vicinity of the origin cannot be ensured. 

On the other hand, \cite{laghrouche21} presents a Higher-Order Sliding Mode Control using barrier functions for systems of arbitrary order. However,  the homogeneous topology induced by the Lyapunov function makes it difficult to prescribe the behavior of the system's states, and the time of convergence to the homogeneous vicinity is not predefined, see the simulation example in \cite{mannydiegocdc}.

The authors of \cite{rodrigues2021} present an adaptive strategy based on the combination of two different controllers: monitoring and barrier functions, providing a predefined upper bound of the settling time to a vicinity of the sliding surface with relative degree one.  Moreover, in \cite{rodrigues2021}, the uncertainties in the control coefficient are not considered.

This paper introduces an adaptive control combining a time-varying feedback gain with barrier function adaptive gain for a perturbed chain of integrators with unknown upper bounds of the perturbations and  unknown control coefficient based on Lyapunov redesign \cite{Gutman1979},\cite{Leitmann1979}. This paper contributes to solving (i)-(iii) with just one controller with a unique switch in the controller's gain:
\begin{itemize}
\item  A PNF type gain is used to reach a prescribed neighborhood of the origin in a predefined time (in the sense of \cite{song2017,gomez2020,aldana2021,chitour2020}), measured in terms of a quadratic Lyapunov function.
The singularity of the time-varying feedback control is avoided because the adaptive barrier function-based control is switched on when the system's solutions reach the interior of the prescribed neighborhood of the origin. 
\item A barrier function-based gain allowing the designer to prescribe the desired neighborhood in terms of the same Lyapunov function, where the trajectories are kept, is presented. Once this neighborhood is reached,  the provided gains ensure that the control signal follows the perturbations.
\item The usage of the topology generated by the same Lyapunov function simplifies the switch between the gains allowing to make it in one step. Moreover, the control signal is continuous and bounded, except at the switching moment.	
\item Finally, proposed results are validated experimentally in the Furuta pendulum system.
\end{itemize}

The organization of the paper is as follows: In Section \ref{sec:PS}, the problem statement is presented. Section \ref{sec:mainresult} contains the control law design and main result of the paper.  Numerical simulations are shown in Section \ref{sec:simex}.  
An experimental result on Furuta's pendulum is given in Section \ref{sec:experiment}. The paper closes with some concluding remarks in Section \ref{sec:conclusion}. Technical proofs are given in Appendix \ref{AppA}.

\textbf{Notation.}  The symbols $\Vert \cdot \Vert$ and $\vert \cdot \vert$ denote the Euclidean norm and the absolute value of a vector and a scalar value, respectively. The $\mathrm{sign}$ function on the real line is defined by $\mathrm{sign}(z) = \tfrac{z}{\vert z \vert}$ for $z \neq 0$ and $\mathrm{sign}(0) = [-1,\:1]$.  For a real square matrix $G$, $\lambda_{\mathrm{min}}(G)$ (resp. $\lambda_{\mathrm{max}}(G)$) denotes the minimum (resp. maximum) eigenvalue of $G$.

\section{Problem Statement}\label{sec:PS}
We address the robust stabilization of the perturbed chain of integrators of the form
\begin{equation}
\begin{aligned}\label{eq:system}
\dot{x}(t)&=J_nx(t)+e_n\left[ b(t)(1+\delta_b(t)) u(t)+f(t) \right] \\
x(0)&=x_0\,,
\end{aligned}
 \end{equation} where $x(t)\in \mathbb{R}^n$ denotes the state of the system, $u(t) \in \mathbb{R}$ is the control input, $(e_i)_{1\leq i \leq n}$ denotes the canonical basis of $\mathbb{R}^n$, $J_n$ denotes the $n$-th Jordan block (i.e., $J_ne_i=e_{i-1}$ for $1\leq i \leq n$ with $e_0=0$) and $b(t)$ is the known part of the control coefficient, without loss of generality assume that $b(t)>\underline{b}>0$. The perturbations $f:\mathbb{R}_+\rightarrow \mathbb{R}$ and $\delta_b(t):\mathbb{R}_+ \rightarrow \mathbb{R}$ are measurable  functions in $t$ where $f(t)$ acquaints for external perturbations to the system and $\delta_b(t)$ is the unknown control coefficient.

 In the sense of quadratic stabilizability \cite{Petersen1987}, without the presence of external perturbations, the stabilization problem of \eqref{eq:system} is readily solved by using a linear state feedback. In fact, exponential stabilization of the closed-loop system is ensured and its trajectories can be confined inside any neighborhood of the origin in a finite time. We assume the following

 \begin{assum}\label{ass1}  For all $t\in \mathbb{R}_{+}$,  there exist unknown constants $M>0$ and $0\leq\varepsilon_b<1$ such that $\vert f(t) \vert \leq M $ and  $\vert \delta_b(t)\vert \leq \varepsilon_{b}$.
\end{assum} Notice, however, that under the above assumption, a linear control law for system \eqref{eq:system} can only ensure ultimate boundedness of the trajectories. Considering the control law 
\begin{equation}\label{eq:nominal}
    u(t)= -\frac{1}{2}\left( \frac{1}{\gamma}+1\right) b^{-1}(t)e_n^TPx(t)\,,
\end{equation} where $\gamma \in \left(0,\frac{1+\sqrt{5}}{2}\right]$ and $P$ is solution of the Algebraic Ricatti Equation (ARE)
\begin{equation}\label{eq:riky}
   J_n^TP + PJ_n -\gamma Pe_ne_n^TP + Q=0, \: Q>0\,,
\end{equation}
    \begin{rem} 
        There exists a value $\gamma^*$ such that the solution of \eqref{eq:riky} exists for all $ \gamma\in \left(0, \gamma^{*}\right)$; see \cite{corless1994}. 
    \end{rem}
\begin{prop}\label{lemma:lineal}
Suppose that Assumption \ref{ass1} is fulfilled. Let $\mu := 2M\lambda_{max}(P)/\left(\theta\lambda_{min}(Q)\right)$ for some $0<\theta<1$, there is $T^*:=T^*(\mu,x_0)$ such that for all $t\geq T^*$, trajectories of system \eqref{eq:system} in closed-loop with \eqref{eq:nominal} satisfy 
\begin{equation}\label{eq:ultimateboundlem1}
    \Vert x(t)\Vert \leq \sqrt{\frac{\lambda_{min}(P)}{\lambda_{max}(P)}}\mu\,.
\end{equation}
\end{prop}
The stabilization problem solved in  Proposition \ref{lemma:lineal} has two main disadvantages:
\begin{itemize}
    \item The ultimate bound depends on the unknown perturbation's upper bound,
    \item The time to reach such ultimate bound  grows without bound with the initial conditions and the perturbation's upper bound.
\end{itemize}
To overcome the above dependency of the perturbation's upper bound on the ultimate bound, one can redesign the control law such that it compensates the perturbations while ensuring exponential stabilization. Following the Lyapunov redesign methodology \cite{gutmancdc76}, one can add a robustifying term to the control law in Proposition \ref{lemma:lineal}. Consider the redesigned control law 
\begin{multline}\label{eq:nominal_redesigned}
    u = -\frac{1}{2}\left( \frac{1}{\gamma}+1\right)b(t)^{-1}e_n^TPx \\
    - \rho b(t)^{-1}\text{sign}\left(\frac{1}{2}\left( \frac{1}{\gamma}+1\right) e_n^TPx\right)
\end{multline} where $P$ is solution to the ARE \eqref{eq:riky} and $0<\gamma \leq \frac{1+\sqrt{5}}{2} $. The following result can be obtained.
\begin{prop}\label{lem:lineal_robust}
    Suppose that Assumption \ref{ass1} is fulfilled. Let $\rho \geq M/(1-\varepsilon_b)$ and given $0\leq \mu^{*}<\Vert x_0 \Vert$, there is $\bar{T}^*=\bar{T}^*(x_0,\mu^{*})$ such that the trajectories of the closed-loop system \eqref{eq:system}-\eqref{eq:nominal_redesigned} satisfy
\begin{equation}\label{eq:ultimateboundlem2}
    \Vert x(t)\Vert \leq \mu^{*}
\end{equation} for all $t\geq \bar{T}^*$
\end{prop}
 However, there are two main concerns that immediately arise:
\begin{itemize}
    \item The gain of the controller depends on the knowledge of the upper bounds of perturbations that are unknown.
    \item The time to reach the predefined neighborhood of the origin grows without bound with the initial condition.
\end{itemize}
In this paper, we adjust the redesigned controller in  \eqref{eq:nominal_redesigned} by introducing adaptive gains to deal with the lack of knowledge of the perturbations' upper bound and incorporating time-varying feedback gains to reach the predefined neighborhood within a predefined reaching time (bounded finite time with a prescribed upper bound). Specifically, for any initial condition, this paper proposes a $u(t)$ that drives the system's trajectories into a prescribed neighborhood of the origin in a predefined time, and after that, the control law ensures that the system's trajectories will never leave that neighborhood despite the presence of perturbations.


\section{Control design and main result}\label{sec:mainresult}
In this section, the control methodology is presented. The proposed approach consists of two phases:
\begin{itemize}
\item First, a control strategy that drives every system's trajectory to a prescribed vicinity of the origin before a predefined time, in the presence of perturbations with an unknown upper bound.
\item A control gain that ensures the ultimate boundedness of the trajectories in the prescribed vicinity of the origin without requiring knowledge of the upper bound of the perturbations.
\end{itemize}
It is important to note that in both phases, the same quadratic Lyapunov function is used.  This Lyapunov function, obtained via the ARE, simplifies the switching between both phases and allows one to deal with the uncertain control coefficient.
\subsection{Predefined time reaching phase}
This subsection provides a predefined reaching phase controller based on a combination of PNF and adaptive gains for the redesigned controller in \eqref{eq:nominal_redesigned}. Consider the reaching phase control law as 
\begin{multline}\label{eq:controlpnf}
u(t) =-{b(t)}^{-1}\left[\frac{1}{2}\left( \frac{1}{\gamma}+1\right) \kappa(t)^n e_n^TP\Omega^{-1}(t)x(t) \right. \\   +\left. \Gamma(t,x(t))\text{sign}\left(\frac{1}{2}\left( \frac{1}{\gamma}+1\right) e_n^TP\Omega^{-1}(t)x(t)\right) \right]
\end{multline}
where  $0<\gamma\leq \tfrac{1+\sqrt{5}}{2}$, $P$ is solution of the ARE  \eqref{eq:riky}, $\Omega(t) = \text{diag}(1,\kappa(t), \ldots, \kappa(t)^{n-1})$ for $\kappa(t) = \frac{1}{\alpha(T-t)}$ a continuous time varying gain and $\Gamma(t,x)$ is the  adaptive gain given by $ \dot{\Gamma}(t,x) = \vert e_n^TP \Omega^{-1}(t)x(t)\vert \kappa(t)^{1-n}$.

It is important to mention that $V(t) = x^T(t)Px(t)$ is a Lyapunov function for the closed-loop of the PNF and the nominal system (i.e.  $u =\gamma {b(t)}^{-1}\kappa(t)^n e_n^TP\Omega^{-1}(t)x(t)$ and $f(t) = 0$ ).  Then,  following \cite{Gutman1979,Leitmann1979} one can add a robustifying term to compensate the perturbation $f(t)$.  Additionally, the  knowledge of the upper bound of the uncertain control coefficient is not needed for the design of the nominal PNF. 
\begin{lem}\label{lem:pnf}
For the closed-loop \eqref{eq:system}-\eqref{eq:controlpnf}, fix $T>0$ and $\varepsilon>0$. If Assumption \ref{ass1} is satisfied and
\begin{equation}\label{eq:alfa}
\alpha < \frac{\lambda_{min}(Q)}{2(n-1)\lambda_{max}(P)}\,,
\end{equation}
 then there exist a first time $\bar{t}(x_0)<T$ such that the solution of the closed-loop system \eqref{eq:system}-\eqref{eq:control} reaches the boundary of the set $ \mathcal{S}_{\frac{\epsilon}{2} }:= \left\lbrace x\in \mathbb{R}^n \:\vert \: V(x)\leq \varepsilon/2 \right\rbrace$. 
\end{lem}

 During the so-called reaching phase, PNF type law increases the proportional feedback until the value allowing compensation of the uncertainties and disturbances, and, consequently, the system's trajectories converge to $\mathcal{S}_{\frac{\epsilon}{2} }$ in a time  $\bar{t}<T$ uniformly for any initial condition and size of perturbations.  Moreover,  the control law is uniformly bounded as shown in Appendix \ref{app:BCL}

  \begin{rem}
Note that for $PJ_n + J_n^TP -\gamma Pe_ne_n^TP = -Q$, the ratio \eqref{eq:alfa} is maximized if $Q = \mathbb{I}$;  see \cite{Patel1978}.
\end{rem} 

\subsection{Barrier function phase }
Once in a predefined neighborhood of the origin, this subsection provides a barrier function phase controller by choosing a positive semidefinite barrier function as gain in the robustifying part of the redesigned controller in \eqref{eq:nominal_redesigned}.Consider the control law
\begin{multline}\label{eq:controlbar}
u(t)=-{b(t)}^{-1}\left[\frac{1}{2}\left( \frac{1}{\gamma}+1\right) e_n^TPx(t) \right. \\ + \left. \frac{V(t)}{\epsilon-V(t)}\text{sign}\left(\frac{1}{2}\left( \frac{1}{\gamma}+1\right) e_n^TPx(t) \right) \right]
\end{multline}
where the function $V(t) = x^T(t)Px(t)$ with $P=P^T>0$ solution of \eqref{eq:riky}.
\begin{lem}\label{lem:barrier}
Suppose that Assumption \ref{ass1} is satisfied. Consider the control law \eqref{eq:controlbar} in closed-loop with system \eqref{eq:system}. For any given $\epsilon>0$ such that $V(x(t_0)) \leq \frac{\epsilon}{2}$, the trajectories of the closed loop system \eqref{eq:system}-\eqref{eq:controlbar} satisfy $V(x(t)) < \epsilon $ for all $t\geq t_0$.
\end{lem}
With the system's trajectories being in the barrier width, the controller barrier function-based law ensures that the system's solution are confined into the set $ \mathcal{S}_{\epsilon }= \left\lbrace x\in \mathbb{R}^n \:\vert \: V(x)< \varepsilon \right\rbrace $  for all $t>t_0$ without the knowledge of the upper bound of the perturbation.
Note that function $V(t) = x^TPx$ is in fact a known Lyapunov function for a \textit{nominal} system (with only $u = -\tfrac{1}{2}\left( \frac{1}{\gamma}+1\right) b(t)^{-1}e_n^TPx(t)$ and  $f(t)=0$), then one can add a robustifying term to compensate $f(t) \neq 0$ in the sense of \cite{Leitmann1979,Gutman1979}. 

Although the idea is similar in \cite{laghrouche21}, the dependency of the barrier function on a homogeneous function $V$ for a homogeneous controller introduces a notable complexity.  The ultimate bound is given in terms of a homogeneous norm, which can be difficult to prescribe and compute.  

For the linear framework given in this work,  prescribing a desired vicinity of the origin results in a straightforward and easy-to-compute methodology.  By the boundedness of $V(t)< \epsilon$, there exist $R>0$ that depends on $\epsilon$ such that $\Vert x \Vert < R$. Then one can design $\epsilon$ such that one can obtain a desired vicinity of the origin. Considering that $ R(\epsilon) = \sqrt{\epsilon/\lambda_{min}(P)}$, for a desired $R>0$, choosing $\epsilon < R^2\lambda_{min}(P)$ will ensure the prescribed region. 


\subsection{Main Result}
The proposed control approach can be summarized as follows:
\begin{equation}\label{eq:control}
\begin{split}
u(t) &=-{b(t)}^{-1}\left[\kappa(t)^n u_0(t) + \Lambda(t,x(t))\text{sign}(u_0(t)) \right],
\end{split}
\end{equation}
where $u_{0}(t)= \frac{1}{2}\left( \frac{1}{\gamma}+1\right) e_n^TP\Omega^{-1}(t)x(t)$, and the switching the function $\kappa(t)$
\begin{equation}\label{eq:TVG}
    \kappa(t) = \begin{cases}
\frac{1}{\alpha(T-t)}    \quad  &\mathrm{if} \,\, t< {T}_1,\, \\
\quad 1 \quad &\mathrm{if} \,\, t \geq {T}_1\,.
    \end{cases}
\end{equation}
and the adaptive gain $\Lambda(t,x)$
\begin{equation*}
    \Lambda(t,x) = \begin{cases}
 \Gamma(t)\,, \: &\mathrm{if} \,\, t< {T}_1,\, \\
 \frac{V}{\epsilon - V}\,,  \: &\mathrm{if}\: t \geq {T}_1\,.
    \end{cases}
\end{equation*}

The time $T_1>0$ denotes the first moment such that $x(T_1)\in \mathcal{S}_{\frac{\epsilon}{2} }$.


\begin{thm}\label{prop1}
Suppose that Assumption \ref{ass1} is satisfied for the closed-loop system \eqref{eq:system}-\eqref{eq:control} and let $P$ be the solution of \eqref{eq:riky}. Given the predefined time $T > 0 $ and a prescribed barrrier width $\epsilon >0$, if $\alpha$ designed as \eqref{eq:alfa},  then the trajectories of the system reach the set $ \mathcal{S}_{\frac{\epsilon}{2} } $  in a time $ T_1 < T$ for all $x_0 \in \mathbb{R}^n$, and will be confined in  $ \mathcal{S}_{\epsilon } $ for all $t\geq T_1$.
\end{thm}
\begin{pf} The proof of this result is made in two steps:
\begin{enumerate}
    \item[A)] First, it is proved that there exists a first moment $0<T_1=\bar{t}(x_0)<T$ such that $x(t)\in \partial  \mathcal{S}_{\frac{\epsilon}{2} }$ with $ \mathcal{S}_{\frac{\epsilon}{2} }=\left\lbrace x(t)\in \mathbb{R}^n\:|\: V(x(t))\leq \varepsilon/2 \right\rbrace$ where $T>0$ is a priori given.
    \item[B)] Secondly, the ultimate boundedness of the trajectories in that region is achieved through the barrier function for all $t > T_1$, independently of the perturbation's upper bound.
\end{enumerate}
Items (A) and (B) are consequences of Lemmas \ref{lem:pnf} and \ref{lem:barrier}, respectively.  \hfill $\square$
 \end{pf}

\begin{rem}
Notice that the structure of \eqref{eq:control} is maintained for both stages and only the gains are switched. However, this only switching causes a discontinuity in the control signal. Therefore, the control law is essentially bounded. It is bounded during the reaching phase (see \ref{app:BCL}) and barrier function phase where $\Gamma(t,x)=\tfrac{V}{\epsilon-V}\leq \tfrac{\sigma_1}{\epsilon-\sigma_1}< \infty $ (see proof of Lemma \ref{lem:barrier}), except at $t=\bar{t}$.
\end{rem}

\section{Simulation Example}\label{sec:simex}
Consider the Torsional spring damper system presented in \cite{Castillo2021}:
 \begin{equation}
  j(t)  \ddot{\theta}(t) + b \dot{\theta}(t) + k\theta = v(t) +\varphi(t)
 \end{equation}
where $j(t)$ is a time-varying inertia, $v(t)$ is an input torque and $\varphi(t)$ is an external disturbance. In this section, the tracking problem of a desired signal $\theta_d$ will be addressed. For that purpose, define the tracking errors: $x_1 = \theta(t) - \theta_d(t)$ and $x_2 = \dot{\theta}(t) - \dot{\theta}_d(t)$ and the nominal control input $v(t)= u(t) +b(x_1 + \theta_d) + k(x_2 + \dot{\theta})$.  Thus, the error dynamics has the following form:
\begin{equation}
\begin{aligned}
\dot{x}_1   &= x_2\\
\dot{x}_2 &=  \frac{1}{j_m}\left[ (1 + \delta_j(t))u(t) \right] + j(t)\left(\ddot{\theta}_d(t) + \varphi(t) \right) \,.
\end{aligned}
\end{equation}
where  $\frac{1}{j(t)} = \frac{1}{j_m}(1 + \delta_j(t) )$ , $j_m$ is the nominal part of the inertia, $\delta_j$ is the uncertainty in the inertia. 

For the simulation, the desired trajectory $\theta_d(t)$ is designed as in \cite{Spong2006}, considering a polynomial trajectory of degree 5 from $t = 0$ to $t = 10$. The desired $\dot{\theta}(0) = 0$ and $\theta_d(10) = 10$, and $\dot{\theta}_d(0)=\dot{\theta}_d(10)=\ddot{\theta}_d(0) = \ddot{\theta}_d(10)=0  $ . Simulation parameters are presented in Table \ref{tab:sim_parameters}. 

\begin{table}[h!]
  \begin{center}
    \caption{Simulation parameters.}
    \label{tab:sim_parameters}
    \begin{tabular}{c|c}
      \textbf{Parameter} & \textbf{Value}\\
      \hline 
      $k$ & 2.3375$\, \text{N}/ m$ \\
      $j$ & 0.2946 $\text{kg}\cdot \text{m}^2 $ \\
      $j_m$ &  $0.0333\, \text{kg}\cdot\text{m}^2$\\
      $b$ &  0.012195 $\text{N}\cdot \text{s}/m$ \\
      $\alpha$ & 0.1\\
      $\gamma$ & 0.1 \\
       $T$ & 2
    \end{tabular}
  \end{center}
\end{table}
Three scenarios are tested in order to show the feasibility of the proposed approach:
    \begin{enumerate}
        \item In the first one, the barrier width is set to $ \epsilon = 1$.
        \item Second one shows the results with $\epsilon = 0.01$.
        \item Finally, the value of the barrier is deacresed to $\epsilon = 1\times 10^4$.
    \end{enumerate}

All simulations were made with $\delta_j(t) = \frac{3}{4}\text{sign}(\sin(t))$ and the external disturbance $\varphi(t) =  \cos(5t) $. The sampling step is set $1\times 10^{-3}$  using Euler integration method and the initial conditions were set to $x(0) = \begin{bmatrix} 5 & 0 \end{bmatrix}^T $. For the control law in Theorem \ref{prop1}, matrix $P$  is obtained as the solution of  Equation \eqref{eq:riky} with $ Q = \mathbb{I}$. The value of  $T$ used for the three scenarios is fixed, as it is presented in Table \ref{tab:sim_parameters}.

\subsection{First scenario}
\begin{figure}[h!]
\centerline{\includegraphics[width=0.55\textwidth]{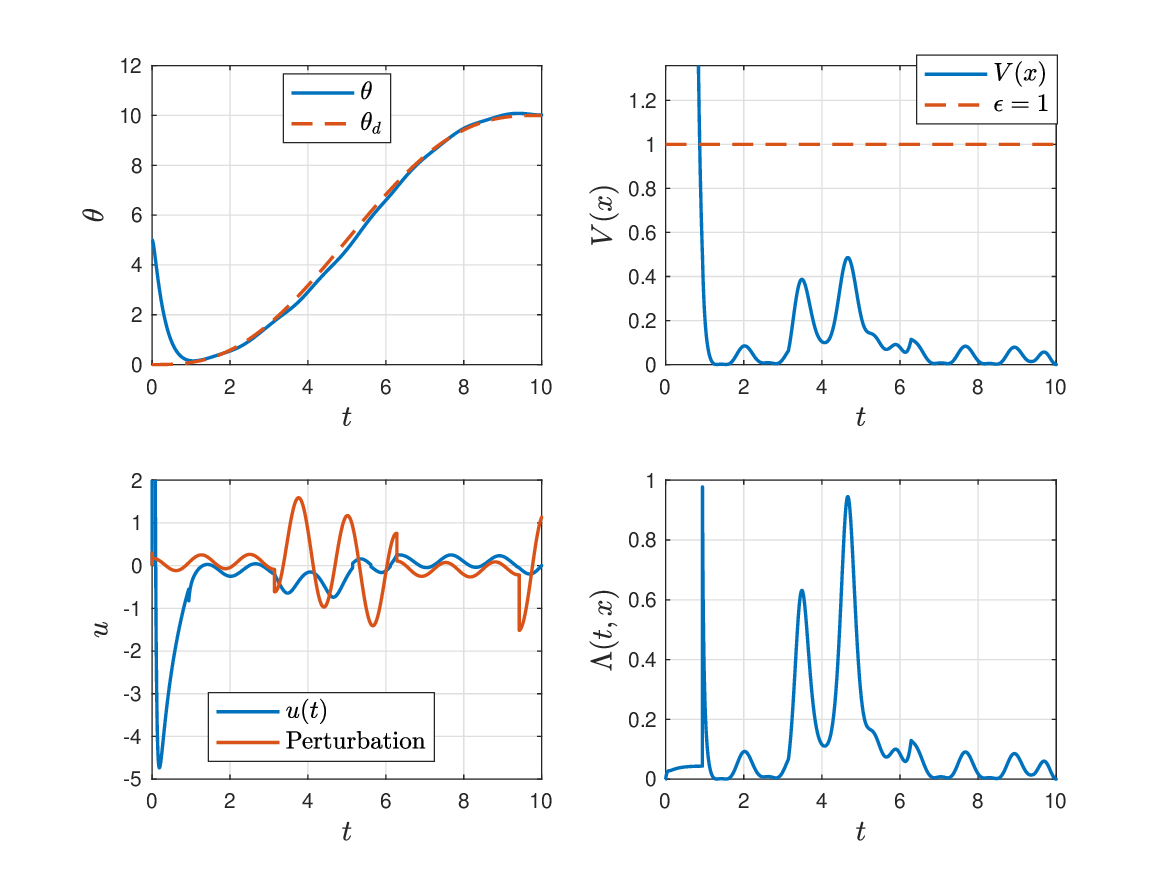}}
\caption{Tracking with $\epsilon = 1$.}
\label{fig:1}
\end{figure}

In Figure \ref{fig:1}  the results of the control law with $\epsilon = 1$ are presented. It can be seen that the tracking is near to the desired angle $\theta_d$, but not exactly the same. As the upper bound of the settling time is designed as $T = 2$, the trajectory converges to a neighborhood of the tracked signal and remain nearby. 
\begin{figure}[h!]
\centerline{\includegraphics[width=0.6\textwidth]{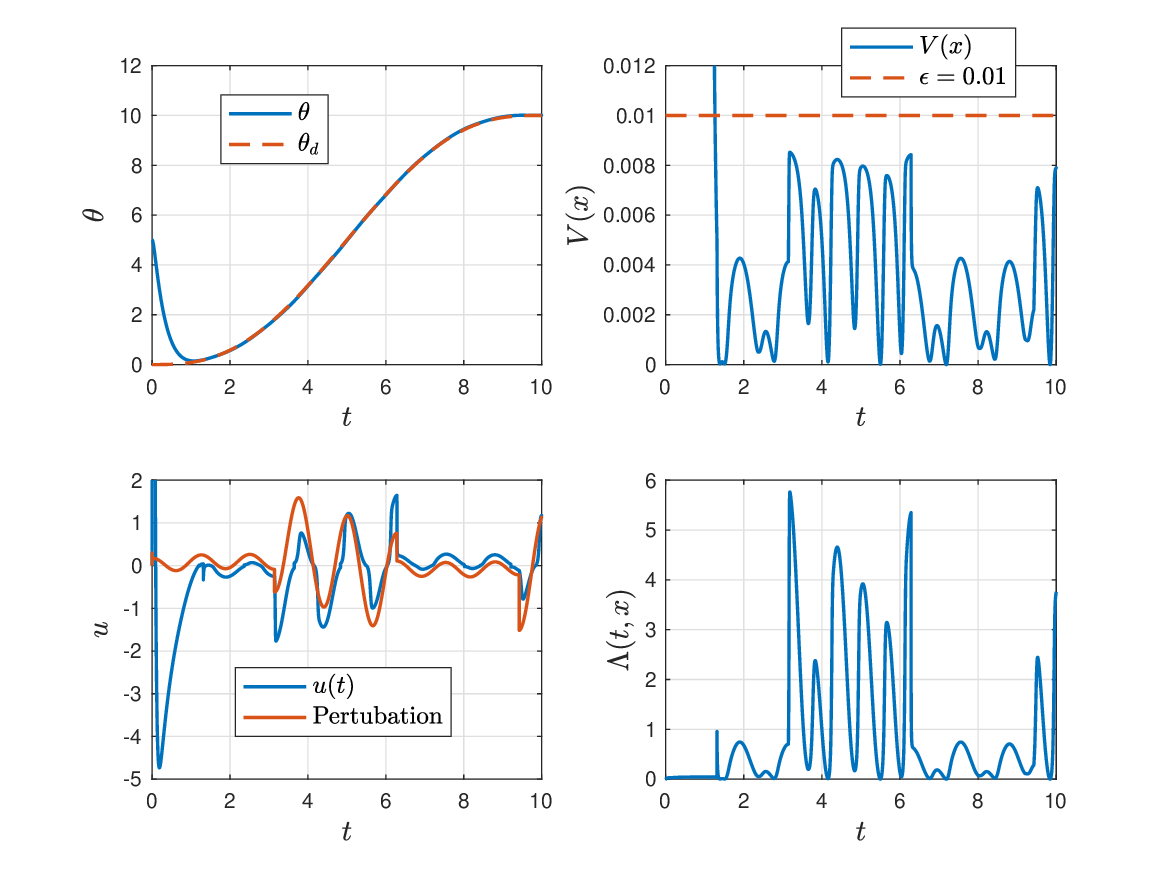}}
\caption{ Tracking with $\epsilon = 1\times 10^{-2}$. }
\label{fig:1e-2}
\end{figure}
\subsection{Second scenario}
For $\epsilon = 0.01$ in Figure \ref{fig:1e-2}, the angle $\theta$ is really close to the tracked desired signal. The control signal $u(t)$ is following the negative of the perturbation. Moreover, the gain $\Lambda(t,x)$ increases considerably with respect to the first scenario. This is a intuitive consequence of the vicinity being smaller.
\begin{figure}[h!]
\centerline{\includegraphics[width=0.6\textwidth]{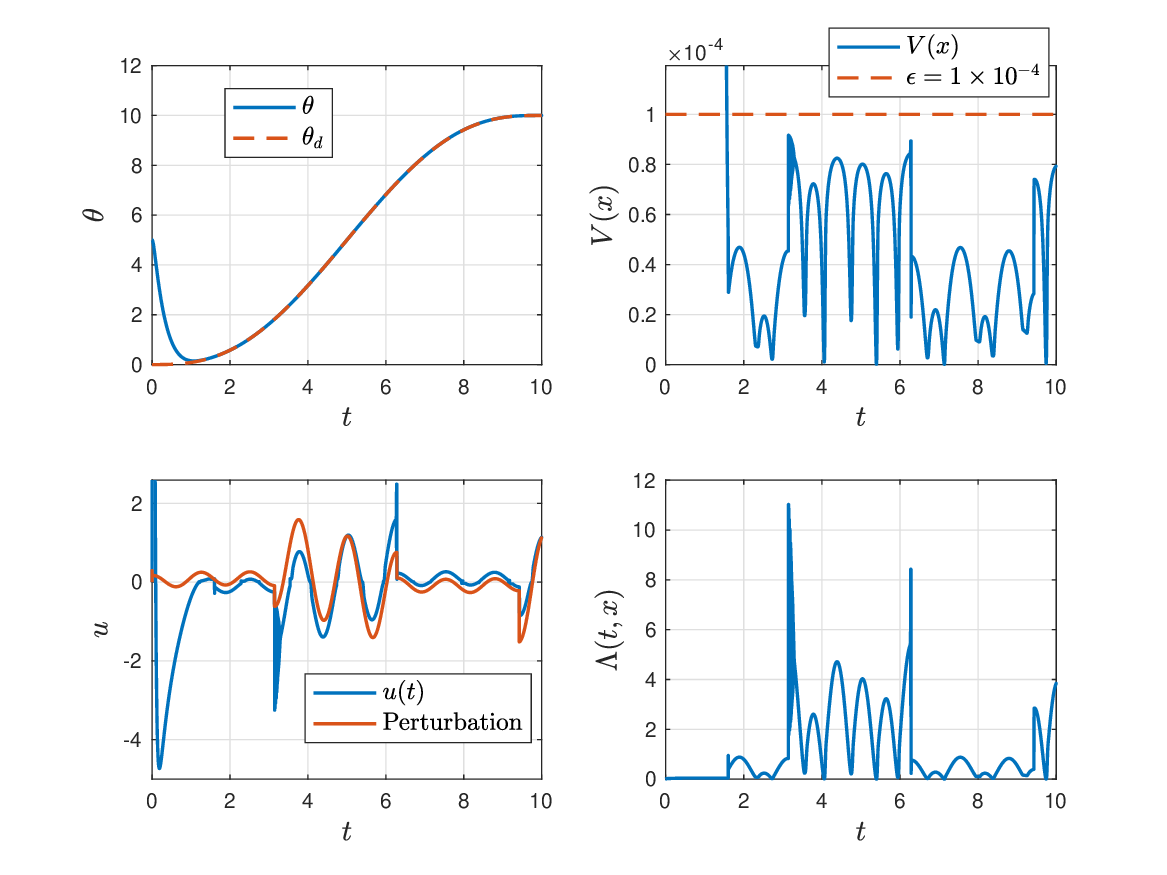}}
\caption{Tracking with $\epsilon = 1\times 10^{-4}$.}
\label{fig:1e-4}
\end{figure}
\subsection{Third scenario}
Finally, Figure \ref{fig:1e-4} presents the tracking selecting $\epsilon = 1\times 10^{-4}$. This small neighborhood of the system's origin is achieved by means of a small barrier width. Nonetheless, a small choice on the barrier width may induce noise in the control signal (see Figure \ref{fig:1e-4}). This illustrates that, although $\epsilon$ can be as small as desired, the sampling step must taken into account in the choice of $\epsilon$.

 \section{Experimental result}\label{sec:experiment}
Consider the Furuta pendulum presented in \cite{m:quanserFuruta} as:
\tiny
\begin{multline*}
\left( m_pL_2^2 + \frac{1}{4}m_pL_p^2\cos^2(\theta_p) + J_r\right)\ddot{\theta}_r - \left( \frac{1}{2}m_pL_pL_r\cos(\theta_p)\right)\ddot{\theta}_p \\
+ \left(\frac{1}{2}m_pL_p^2\sin(\theta_p)\cos(\theta_p)\right)\dot{\theta}_r\dot{\theta}_p + \left(\frac{1}{2}m_pL_pL_r\sin(\theta_p)\right)\dot{\theta}_p^2 = \tau \\
-\frac{1}{2}m_pL_pL_r\cos(\theta_p)\ddot{\theta_r} + \left(J_p + \frac{1}{4}m_pL_p^2\right) \ddot{\theta}_p - \frac{1}{4}m_pL_p^2\cos(\theta_p)\sin(\theta_p) \dot{\theta}_r^2\\
-\frac{1}{2}m_pL_pg\sin(\theta_p) = 0
\end{multline*}
\normalsize
where $\theta_r$ and $\theta_p$ are the angles for the arm and the pendulum, respectively, the parameter $m_p$ is the mass of the pendulum, $L_p$ and $L_r$ are the lengths of the pendulums and arm and $J_r$ is the inertia of the arm. One can compute the torque-voltage (input to the motor $V_m$) conversion as:
\begin{equation}\label{eq:conversion}
 \tau = \frac{\eta_gK_g\eta_mk_t\left(V_m - K_gk_m\dot{\theta}_r  \right)}{R_m}\,.
\end{equation}
where $\eta_g, K_g, k_t, R_m, \eta_g$ are motor parameters. For $z_1=\theta_r,$ $z_2=\theta_p$, $z_3=\dot{\theta}_r$ and $z_4 =\dot{\theta}_p$, linearising around the point $z^T = \begin{bmatrix}0 & 0 & 0 &0 \end{bmatrix}$ with $z$ being a vector of the elements $z_i$, it yields to a form $\dot{z} = Az + B\tau$ with matrices
\begin{align*}
A &= \frac{1}{J_T} \begin{bmatrix}
0 & 0 & 1 & 0 \\ 0 & 0 & 0 & 1 \\ 0 & \frac{1}{4}m_pL_p^2L_rg & 0 & 0 \\ 0 & \frac{1}{2}m_pL_pg(J_r + m_pL_r^2) & 0 & 0
\end{bmatrix},\, \\ B &= \frac{1}{J_T}\begin{bmatrix}
0 \\ 0 \\ J_p +  \frac{1}{4}m_pL_p^2 \\ \frac{1}{2}m_pL_pL_r
\end{bmatrix}
\end{align*}
and $J_T = J_pm_pL_r^2 + J_rJ_p + \frac{1}{4}J_rm_pL_p^2$. One can then find a matrix transformation $W = \begin{bmatrix} B & AB & A^2B & A^3B \end{bmatrix}H_k$, where $H_k$ is a Henkel matrix with the elements of the characteristic polinomial of $A$, such that the system is led to the controller form, then taking $x = Wz$, the system will be in the required form if $\tau = -\tau_n x_3 + u$, where $\tau_n = 0.1112$ is a constant depending on the parameters of the plant.  The parameters proposed for the experiment where is solution of \eqref{eq:riky} with $\gamma = 0.45$ and $Q=\mathbb{I}$ and to satisfied condition \eqref{eq:alfa}, $\alpha = 0.002$. The integrator intial condition $\Gamma(0) = 0$ and the sampling step provide of $ 1 \, \text{ms}$, the following two scenarios where tested in the Furuta pendulum system by \textit{Quanser Inc}\textregistered. Note that the input of the pendulum system is saturated from $u\in \left[-10,10\right]\, \text{volts}$ to the motor, with the relation given by \eqref{eq:conversion}.
\subsection{For different upperbound of the settling time}
\begin{figure}[h!]
\centerline{\includegraphics[width=0.5\textwidth]{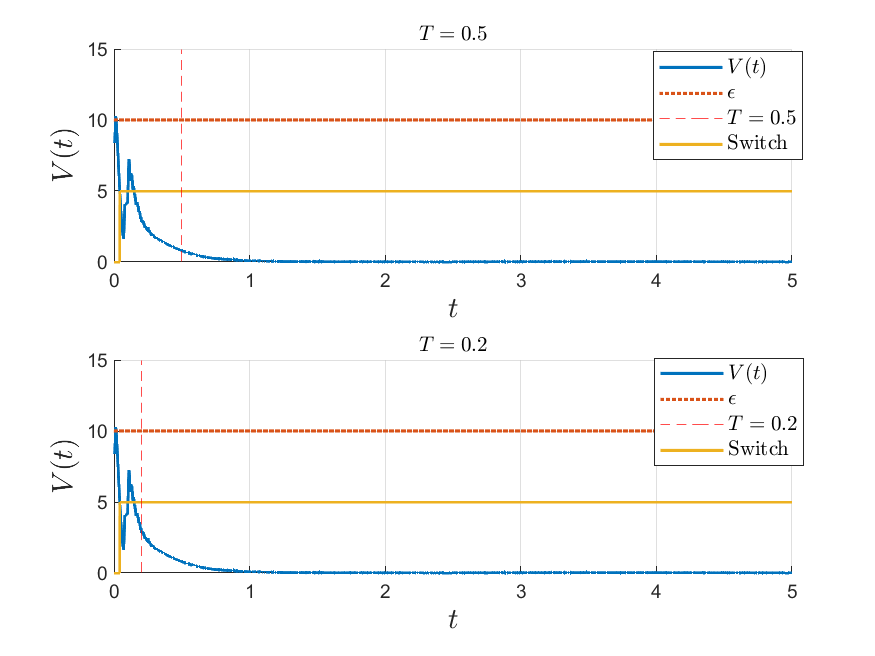}}
\caption{Values of $V(t)$ for different values of $T$ starting from $\theta_p(0) = 0.5$.}
\label{fig:VT}
\end{figure}

In this subsection, the initial condition is the same for both cases $\theta_p(0) = 0.5$, setting $\epsilon = 10$ and the upper bound of the settling time is selected $T=0.5$ and $T= 0.2$. It can be seen from Figure \ref{fig:VT} that both of the settling time bounds are maintained to the desired vicinity of the origin by means of the function $V(t)$. 
 
  \begin{figure}[h!]
\centerline{\includegraphics[width=0.5\textwidth]{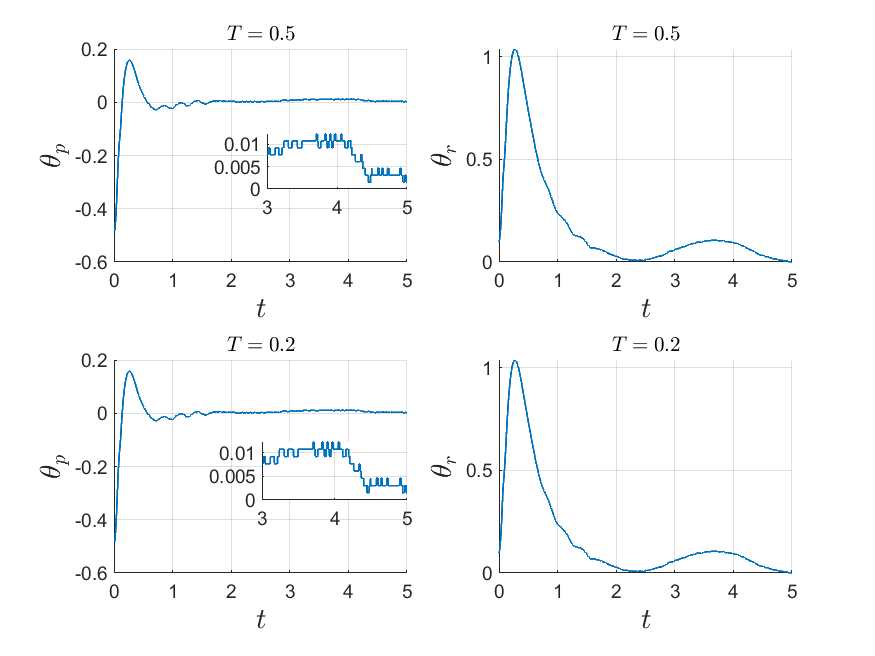}}
\caption{Angles for different values of $T$ starting from $\theta_p(0) = 0.5$.}
\label{fig:thetasT}
\end{figure}

  \begin{figure}[h!]
\centerline{\includegraphics[width=0.5\textwidth]{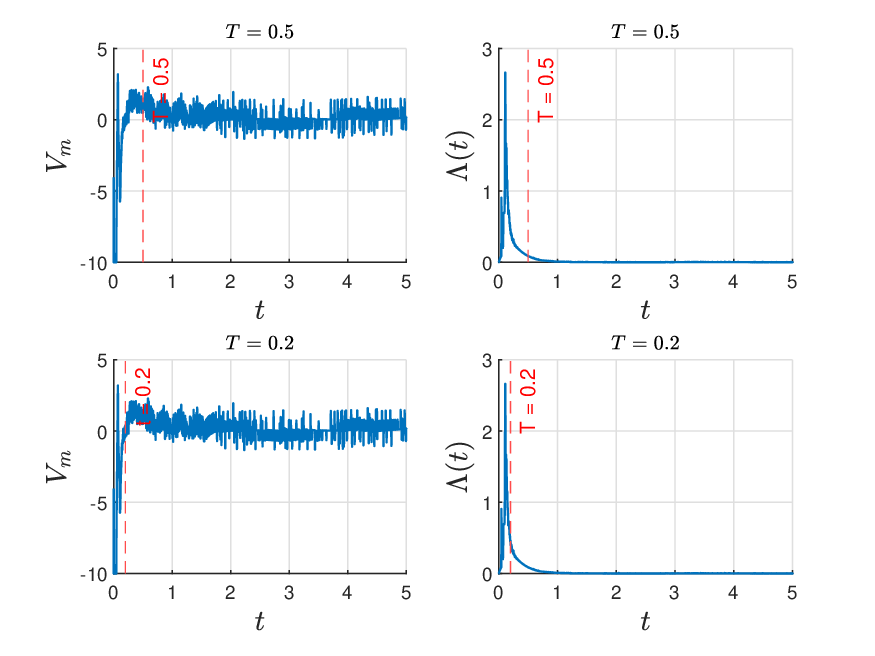}}
\caption{control signal and gain $\Lambda(t)$ for different values of $T$ starting from $\theta_p(0) = 0.5$.}
\label{fig:usT}
\end{figure}

 \subsection{For different prescribed neighborhood}
\begin{figure}[h!]
\centerline{\includegraphics[width=0.6\textwidth]{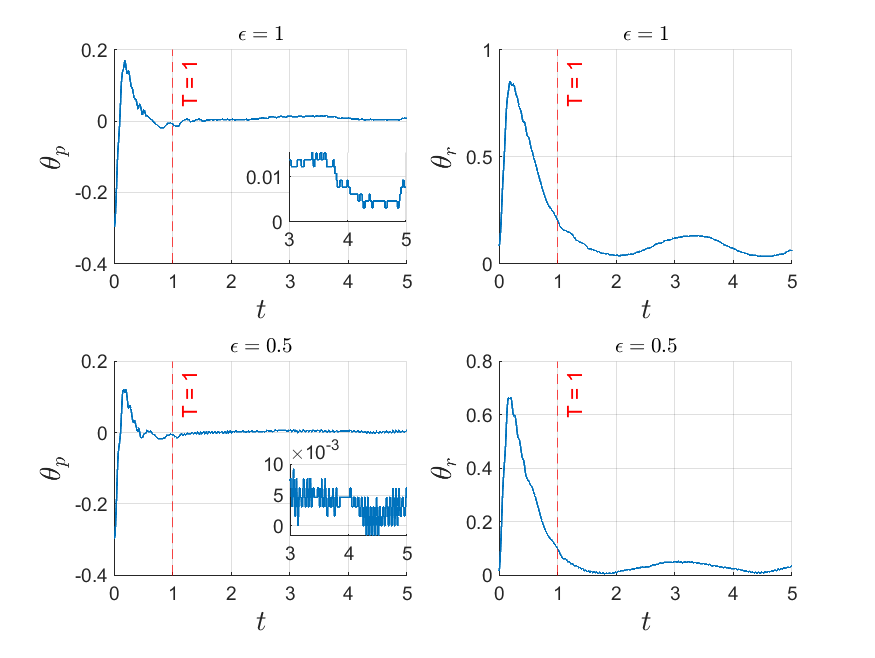}}
\caption{Angles for different values of $\epsilon$ with $\theta_p(0)=0.3$ and $ T = 1$.}
\label{fig:thetaseps}
\end{figure}

\begin{figure}[h!]
\centerline{\includegraphics[width=0.5\textwidth]{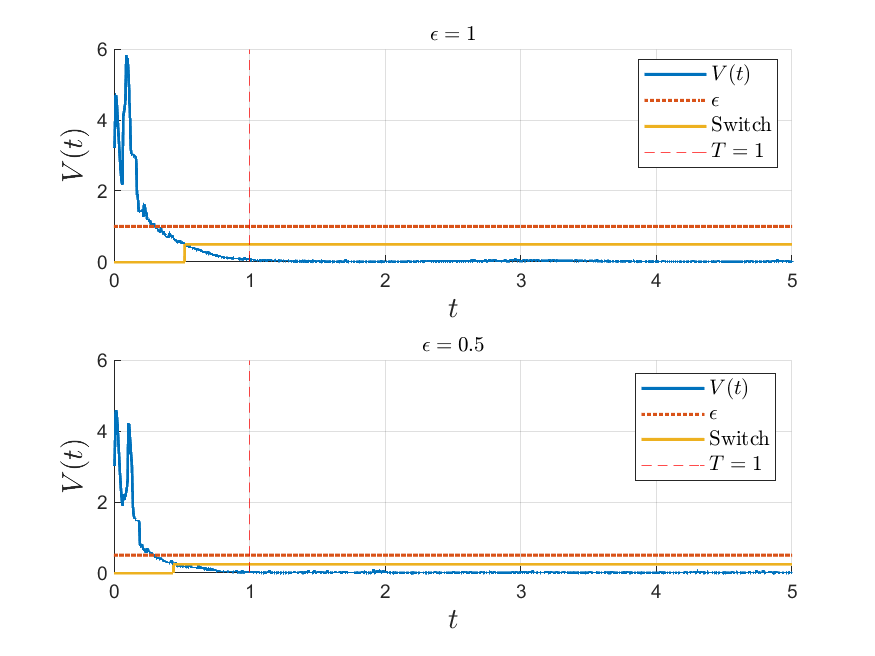}}
\caption{Values of $V(t)$ for different $\epsilon$ with $\theta_p(0)=0.3$ and $ T = 1$.}
\label{fig:Veps}
\end{figure}

Additionally, initiating from $\theta_p(0) = 0.3$ and setting $T = 1$, the next demonstration is the capability of the approach in order to prescribe different values for the $\epsilon$-vicinity of the origin of $V =0$. It can be seen from Figure \ref{fig:thetaseps} that for the $\epsilon = 0.5$, the vicinity of the origin for the angle of the pendulum $\theta_p$ is smaller, near to $5\times 10^{-3}\,  \text{rad}$, than the angle for $\epsilon = 1$, that is around $0.015\, \text{rad}$.
 As well as subsection V.A, from Figure \ref{fig:Veps}, it can be seen that the states converge to the prescribed $\epsilon$-vicinity of the origin of $V(t) = 0$ before the predefined time $T = 1$, and the moment of the switching is represented by the yellow line, where $V = \epsilon/2$. The video of the experiments can be found in the following link \url{https://www.youtube.com/shorts/1mOcnIYYLLs}.
 
\begin{figure}[h!]
\centerline{\includegraphics[width=0.5\textwidth]{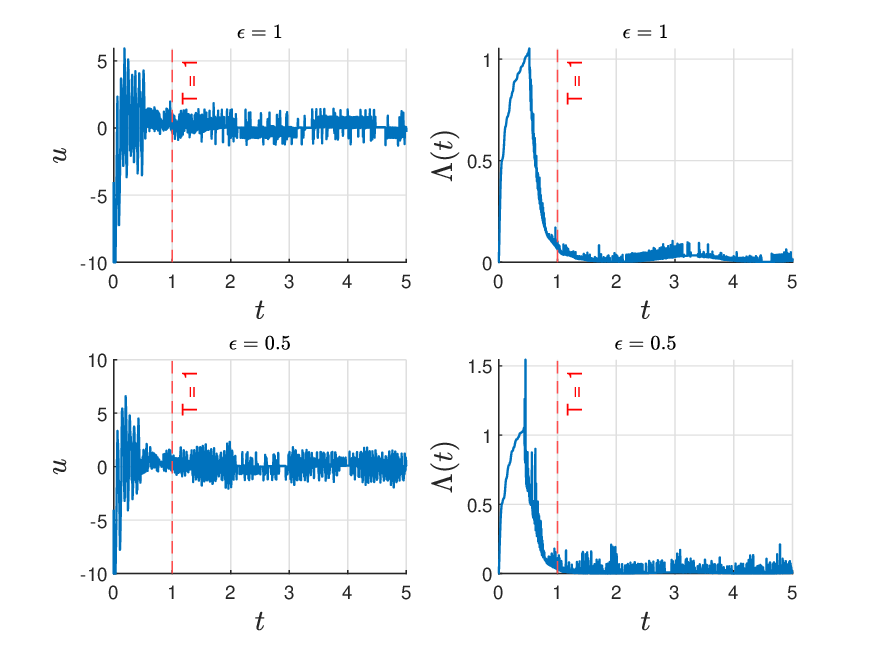}}
\caption{Control signal and gain $\Lambda(t)$ for different values of $\epsilon$ with $\theta_p(0)=0.3$ and $ T = 1$.}
\label{fig:useps}
\end{figure}
 Note also that from Figure \ref{fig:Veps}, it can be seen that the transitory behavior from the experiment with $\epsilon = 0.5$ in the sense of the Lyapunov function is considerably better than the one with $\epsilon =1$, where the overshoot in $V(t)$ is near $6$.

 \section{Conclusion}\label{sec:conclusion}
A Lyapunov redesign methodology is proposed to confine a trajectory of the system modeled by the perturbed chain of integrators in the prescribed vicinity of origin in a predefined time, even for the case when the upper bound of perturbation exists but is unknown. The efficacy of the proposed approach is illustrated through the simulations for the spring-mass model and experiments with the Furuta pendulum.
 
 \section*{Acknowledgments}                               
The authors are grateful for the financial support of Programa de Becas Posdoctorales DGAPA-UNAM, CONACyT (Consejo Nacional de Ciencia y Tecnolog\'ia): Project 282013, and CVU 833748; PAPIIT-UNAM (Programa de Apoyo a Proyectos de Investigaci\'on e Innovaci\'on Tecnol\'ogica): Project IN106622. 

\bibliographystyle{plain}        
\bibliography{bibliog}           

\appendix
\section{Technical proofs}\label{AppA}
\subsection{Proof of Proposition \ref{lemma:lineal}}\label{proof:Lemma1}
For the closed-loop system \eqref{eq:system}-\eqref{eq:nominal} of the form 
\begin{equation}\label{eq:syslem1}
    \dot{x}(t) = J_n x(t) -\frac{1}{2}\left( \frac{1}{\gamma}+1\right)(1+\delta_b(t))e_ne_n^TPx(t)+e_nf(t)\,,
\end{equation}
consider the Lyapunov function candidate $V(x) = x(t)^TPx(t)$, its time  derivative along the trajectories of \eqref{eq:syslem1} yields to
\begin{equation}
    \begin{split}
        \dot{V}(x)&= x(t)^T\left[ J_n^TP + PJ_n\right]x(t) + 2f(t)e^TPx(t)\\
        &-\left( \frac{1}{\gamma}+1\right)(1+\delta_b(t))x^T(t)Pe_ne_n^TP x(t)
    \end{split}
\end{equation}
             Using the bound $\vert \delta_b(t) \vert\leq \varepsilon_b <1$ from Assumption 1 and 3, the following upper bound of $\dot V(x)$ is found:
            \begin{equation}
            \begin{split}
            \dot{V}(x) &\leq x(t)^T\left[PJ_n+J_n^TP-\left( \frac{1}{\gamma}+1\right)\left( 1-\varepsilon_b\right) Pe_ne_n^TP\right]x(t)\\ & +2x(t)^TPe_nf(t) \,.
            \end{split}
            \end{equation}
Let us analyze when the following will happen:
 	            \begin{equation}
            \begin{split}
 			\left( \frac{1}{\gamma}+1\right)\left( 1-\varepsilon_b\right) = \gamma\,,\,\,
 			\Rightarrow\quad {\varepsilon}_b = 1- \frac{\gamma^2}{1+\gamma}
            \end{split}
            \end{equation}     
            where we can parameterize the bound $\varepsilon_b$ by choosing $\gamma$ in order to compensate the effect of the uncertain coefficient.  Note that this bound will hold as long as $\varepsilon_b<1$. On the other hand, if there does not exist the effect of the perturbation, $\gamma$ can be chosen as the solution of $1+\gamma-\gamma^2=0$, which is $\gamma = \frac{1+\sqrt{5}}{2}$.  If that is the case, then for $\gamma \in  \left(0,\frac{1+\sqrt{5}}{2}\right]$,  the following upper bound is fulfilled
        \begin{align*}
            \dot{V}(x) &\leq x(t)^T\left(PJ_n+J_n^TP- \gamma Pe_ne_n^TP\right)x(t)\\ &+2x(t)^TPe_nf(t)\\
            &=-x(t)^TQx(t)+2x(t)^TPe_nf(t)
        \end{align*}
where we used the equality in \eqref{eq:riky}. By using the upper bound of $\vert f(t) \vert \leq M$ in Assumption \ref{ass1}, the following bound can be obtained by applying the Cauchy-Schwarz and Rayleigh-Ritz inequalities
\begin{equation}
    \begin{aligned}
    \dot{V}(x)\leq-\lambda_{\min}(Q)\Vert x(t) \Vert^2+2M\vert x(t)^TPe_n \vert
    \end{aligned}
\end{equation} Notice that the second term in the right-hand side of the above inequality can be further amplified by using the Cholesky decomposition \cite{horn1990}. There exist $R$ such that $P=RR^T$ and $\vert x(t)^TRR^Te_n\vert\leq \| x(t)^TR\| \| R^Te_n\|$. From the fact that $\Vert x(t)^TR \Vert = V(x)^{1/2}\leq \lambda_{\max}^{1/2}(P)\Vert x(t) \Vert
$ and $\Vert R^T e_n \Vert\leq V(e_n)^{1/2}\leq \lambda_{\max}^{1/2}(P)\|e_n\|=\lambda_{\max}^{1/2}(P)$. It follows that, 
\begin{equation}
    \begin{aligned}
    \dot{V}(x)\leq -\lambda_{\min}(Q)\Vert x(t) \Vert^2+2M\lambda_{\max}(P)\Vert x(t) \Vert
    \end{aligned}
\end{equation}
Hence, the foregoing inequality can be rewritten as 
\begin{equation*}
    \begin{gathered}
\dot{V}(x) \leq -(1-\theta)\lambda_{min}(Q)\Vert x(t)\Vert^2\,,  \forall\, \Vert x(t)\Vert \geq \mu
    \end{gathered}
\end{equation*} where $0<\theta<1$ and $\mu=\frac{2M\lambda_{max}(P)}{\theta \lambda_{min}(Q)}$.  It follows from Theorem 4.18 in \cite{Khalil2001} that  there is $T^*:=T^*(\mu,x_0)$ such that the solution of the closed-loop system \eqref{eq:syslem1} is uniformly bounded, for all $t\geq T^*$ and an initial state $x(0)$ with ultimate bound given by \eqref{eq:ultimateboundlem1}. \hfill $\square$
\subsection{Proof of Proposition \ref{lem:lineal_robust}}\label{proof:Lemma2}
For the closed-loop system \eqref{eq:system}-\eqref{eq:nominal_redesigned} of the form
\begin{multline}\label{eq:system2cl}
    \dot{x}(t)=J_nx(t)-\frac{1}{2}\left( \frac{1}{\gamma}+1\right)(1+\delta_b(t))e_ne_n^TPx(t)\\-e_n\rho(t,x)(1+\delta_b(t))\mathrm{sign}\left(\frac{1}{2}\left( \frac{1}{\gamma}+1\right) e_n^TPx(t)\right)\\
    +e_nf(t)
     \end{multline} consider the Lyapunov function candidate $V(x)=x(t)^TPx(t)$. Using similar arguments as in the proof Proposition \ref{lemma:lineal}, the time derivative of $V(x)$ along the trajectories of system \eqref{eq:system2cl} accepts the following upper bound
\begin{multline}
        \dot{V}(x)\leq -\lambda_{\min}(Q) \Vert x(t)\Vert^2+2x^T(t)Pe_nf(t)-2x^T(t)Pe_n \\ \times \left[\rho(t,x)(1+\delta_b(t))\text{sign}\left(\frac{1}{2}\left( \frac{1}{\gamma}+1\right) e_n^TPx(t)\right)\right]
\end{multline}
 by using the upper bounds in Assumption \ref{ass1} and choosing $\rho(t,x)\geq M/(1-\varepsilon_b)$, the above expression can be rewritten as follows
\begin{equation}
\begin{aligned}
    \dot{V}(x)\leq& -\lambda_{\min}(Q) \Vert x(t) \Vert^2-(1-\varepsilon_b)\rho(t,x) \vert 2e_n^TPx(t)\vert  \\ 
    & + M \vert 2e_n^TPx(t)\vert \\ 
    & \leq -\lambda_{\min}(Q) \Vert x(t) \Vert^2 \leq - \frac{\lambda_{\min}(Q)}{\lambda_{\max}(P)}V(x)
\end{aligned}
\end{equation} where we applied the Cauchy-Schwarz and Rayleigh-Ritz inequalities. The above inequality accepts the solution $V(x)\leq \exp{(-\tfrac{\lambda_{\min}(Q)}{\lambda_{\max}(P)}t)}V(x_0)$, equivalently, $\Vert x(t) \Vert \leq \sqrt{\tfrac{\lambda_{\max}(P)}{\lambda_{\min}(P)}}\exp{(-\tfrac{\lambda_{\min}(Q)}{2\lambda_{\max}(P)}t)}\Vert x_0 \Vert$. Hence, the time $\bar{T}^*$ required for a trajectory starting at $x_0$ to reach the value $0<\mu^{*}<\Vert x_0 \Vert$ is given by $$\bar{T}^*=\tfrac{2\lambda_{\max}(P)}{\lambda_{\min}(Q)}\ln{\left( \sqrt{\tfrac{\lambda_{\max}(P)}{\lambda_{\min}(P)}} \tfrac{\Vert x_0 \Vert}{\mu^*}\right)}.$$
\subsection{Proof of Lemma \ref{lem:pnf}}\label{app:PNF}
If $x(t_0)\in  \mathcal{S}_{\frac{\epsilon}{2} }$, set $\bar{t}(x_0)=0$ and the proof is done. Assume that this is not the case, \textit{i.e.},  $x(t_0)\in  \mathcal{S}_{\frac{\epsilon}{2} }^{c}$. By using the time varying coordinate transformation in \cite{gomez2020}, $x=\Omega(t)y$,
\begin{equation}\label{eq:Tomega}
\Omega(t)=\mathrm{diag}(1,\kappa(t),\kappa(t)^2,\ldots,\kappa(t)^{n-1})
\end{equation} where $\kappa(t)=\frac{1}{\alpha(T-t)}$, $\alpha\in (0,1)$, the perturbed chain of integrators \eqref{eq:system} yields to
\begin{equation*}
\begin{aligned}
\dot{y}_i&=\alpha(1-i)\kappa(t)y_i+\kappa(t)y_{i+1}, \: 1\leq i \leq n -1\\
\dot{y}_n&=\alpha(1-n)\kappa(t)y_n+\kappa(t)^{1-n}\left[b(t)(1+\delta_b(t))u(t)+f(t) \right]
\end{aligned}
\end{equation*} Now take the time scaling $t(\tau)=T(1-\mathrm{e}^{-\alpha \tau})$, by using the chain rule $\frac{d}{dt}y_i\frac{dt}{d\tau}$, where $\frac{dt}{d\tau}=\alpha T\mathrm{e}^{-\alpha \tau}=:\kappa(\tau)^{-1}$, one has
\begin{equation*}
\begin{aligned}
y^{'}_i&=\dot{y}_i\kappa(\tau)^{-1}=-\alpha(i-1)y_i+y_{i+1}, \: 1\leq i \leq n\\
y^{'}_n&=-\alpha(n-1)y_n+\kappa(\tau)^{-n}\left[ b(\tau)(1+\delta_b(\tau)) u(\tau)+f(\tau)\right]
\end{aligned}
\end{equation*} or equivalently in compact form
\begin{equation}\label{eq:system2}
y^{'}(\tau)=Ay(\tau)+e_n\kappa(\tau)^{-n}\left[b(\tau)(1+\delta_b(\tau))u(\tau)+f(\tau) \right]
\end{equation} where $A=J_n+\alpha D_{\alpha}$, and $D_{\alpha}=\mathrm{diag}(0,\:-1,\:\ldots,-(n-2),-(n-1))$.
After applying the time scaling, the first part of the control law in \eqref{eq:control} is given by 
 \begin{equation}\label{eq:pnf}
     \begin{split}
u(\tau)=&- \frac{1}{2}\left( \frac{1}{\gamma}+1\right) b(\tau)^{-1} e_n^TPy(\tau)\kappa(\tau)^n\\
&-\Gamma(\tau)\mathrm{sign}\left(\frac{1}{2}\left( \frac{1}{\gamma}+1\right) e_n^TPy(\tau)\right)
\end{split}
\end{equation} with the adaptive gain \begin{equation}\label{eq:adaptau}
    \Gamma'(\tau) = \vert e_n^TPy(\tau) \vert \kappa(\tau)^{-n}\,,
\end{equation} the closed-loop system has the form:
\begin{multline}\label{eq:clsys_scaledtime}
    y'(\tau)=Ay(\tau)-\frac{1}{2}\left( \frac{1}{\gamma}+1\right) (1+\delta_b(\tau))e_ne_n^TPy(\tau)\\
    -\kappa(\tau)^{-n}\left( \Gamma(\tau)(1+\delta_b(\tau))\mathrm{sign}(\tilde{\gamma} e_n^TPy(\tau))-f(\tau) \right)e_n
\end{multline}
where $\tilde{\gamma} = \frac{1}{2}\left( \frac{1}{\gamma}+1\right)$. Consider the Lyapunov $\bar{V}(\tau)=V_1(\tau)+V_2(\tau)$ where
\begin{equation}
    V_1(\tau)=y(\tau)Py(\tau),\:V_2(\tau)=(1-\varepsilon_b)(\Gamma(\tau)-\Gamma^{*})^2
\end{equation} and $\Gamma^{*}=M/(1-\varepsilon_b)$.

\begin{itemize}
\item For $V_1(\tau)$, its time derivative along the trajectories of \eqref{eq:clsys_scaledtime} yields to
\end{itemize}
\begin{equation*}
\begin{aligned}
&V_1'(\tau) = y^T(\tau)Py'(\tau) + (y(\tau)')^TPy(\tau)\\ 
&\leq y^T(\tau)\left(PA+A^TP \right)y(\tau)-2\kappa(\tau)^{-n}\left[ \Gamma(\tau)(1-\varepsilon_b)\right. \\ &\left.-M\right]\vert e_n^TPy(\tau)\vert-\left( \frac{1}{\gamma}+1\right) (1+\delta_b(\tau))y(\tau)^TPe_ne_n^TPy(\tau)
\end{aligned}
\end{equation*} where we used the bounds for the perturbation terms in Assumption \ref{ass1}. Similarly than in the proof of Proposition \ref{lemma:lineal}, the following upper bound is satisfied
\begin{align*}
&V_1'(\tau) \leq y(\tau)^T\left(PA+A^TP-\gamma Pe_ne_n^TP\right)y(\tau)\\&-2\kappa(\tau)^{-n}(1-\varepsilon_b)\left(\Gamma(\tau)-\Gamma^{*}\right)\vert e_n^TPy(\tau)\vert\,.
\end{align*}
 Finally,  since $1-\delta_b(\tau)^2>0$ by Assumption \ref{ass1} and taking the solution of \eqref{eq:riky}
\begin{equation*}
\begin{aligned}
V_1'(\tau) &\leq- \lambda_{min}(Q)\Vert y(\tau)\Vert^2 +2\alpha y^TD_{\alpha}Py\\& -2\kappa(\tau)^{-n}(1-\varepsilon_b)\left(\Gamma(\tau)-\Gamma^{*}\right)\vert e_n^TPy(\tau)\vert \\
&\leq -\lambda_{min}(Q)\Vert y(\tau) \Vert^2 + 2\alpha(n-1)\lambda_{max}(P)\Vert y(\tau) \Vert^2\\& -2\kappa(\tau)^{-n}(1-\varepsilon_b)\left(\Gamma(\tau)-\Gamma^{*}\right)\vert e_n^TPy(\tau)\vert 
\end{aligned}
\end{equation*}
and therefore,  if $\alpha$ is selected as \eqref{eq:alfa}, then there exist $\bar{\alpha}>0$ such that
\begin{multline}\label{eq:V1tilde}
V_1'(\tau) \leq -\bar{\alpha}\Vert y(\tau) \Vert^2\\  -2\kappa(\tau)^{-n}(1-\varepsilon_b)\left(\Gamma(\tau)-\Gamma^{*}\right)\vert e_n^TPy(\tau)\vert 
\end{multline}

 \begin{itemize}
\item For $V_2(\tau)$, its time derivative along the solution of \eqref{eq:adaptau} yields to
\end{itemize}
\begin{align}\label{eq:V2tilde}
V_2'(\tau) &\leq 2\left( 1-\varepsilon_b\right)\Gamma'(\tau)\left[ \Gamma(\tau) - \Gamma^{*}\right]\\
&=2\kappa(\tau)^{-n}(1-\varepsilon_b)\left(\Gamma(\tau)-\Gamma^{*}\right)\vert e_n^TPy(\tau)\vert\,,\nonumber
\end{align}

From the upper bounds in \eqref{eq:V1tilde} and \eqref{eq:V2tilde}, then $\bar{V} \leq-\bar{\alpha}\Vert y\Vert^2 =-W(y(\tau)) \leq 0$. This implies that $\bar{V}(\tau)\leq \bar{V}(0)$, hence $y(\tau)$ and $\Gamma(\tau)$ are bounded. Noticing that $\int_{0}^{\infty}W(y(s))\mathrm{d}s\leq \bar{V}_0-\lim_{\tau \rightarrow \infty}\bar{V}(\tau)<\infty$. On the other hand, from the continuity of $W(y(\tau))$, and uniform continuity (this follows by noticing that the time derivative of $y(\tau)$ is bounded, using Assumption \ref{ass1}, each term in the time derivative \eqref{eq:clsys_scaledtime} is bounded) and boundedness of $y(\tau)$, then $W(y(\tau)$ is uniformly continuous. By using Barbalat's Lemma \cite{Khalil2001}, then $W(y(\tau))\rightarrow 0$ as $\tau \rightarrow \infty$, this implies that $y(\tau)$ reaches the set $\left\lbrace \Vert (y(\tau))\Vert=0 \right\rbrace$ as $\tau$ grows unbounded. Hence, by continuity, there exists a time $\bar{\tau}(y(0))<\infty$ such that the solution reaches any neighborhood of the zero. Equivalently, after using the inverse time scale transformation $\tau(t)=-\tfrac{1}{\alpha}\mathrm{ln}(1-\tfrac{t}{T})$, the solution $x(t)=\Omega(t)^{-1}y(t)$ reaches any neighborhood of zero in a time $\bar{t}(x_0)=\lim_{\tau \rightarrow \bar{\tau}}T_c(1-e^{-\tau})<T$. The proof is complete after setting the time $\bar{t}$ as the time when $x(t)$ reaches the boundary of the set $ \mathcal{S}_{\frac{\epsilon}{2} }$. 
\subsection{Boundedness of control law during reaching phase} \label{app:BCL}
One can proceed to prove that the control law \eqref{eq:control} is bounded by noticing that it accepts the following upper bound
\begin{align}
    \vert u(t) \vert \leq \underline{b}^{-1}[\kappa(t)^n\vert u_0(t) \vert + \Gamma(t)]
\end{align} Then, we proceed to prove that each summand on the right hand side in the above expression is bounded in the following two steps. 

\noindent \textbf{Step 1.} For the first summand, consider the closed-loop system in the scaled-time \eqref{eq:clsys_scaledtime}, and the Lyapunov-like function $V_1(\tau)=y^T(\tau)Py(\tau)$. The time derivative of $V_1(\tau)$ along the trajectories of \eqref{eq:clsys_scaledtime} satisfies the upper bound \eqref{eq:V1tilde}, which can be further overestimated as follows
    \[
        V_1^{'}\leq -\bar{\alpha}\Vert y(\tau) \Vert^2-2\kappa(\tau)^{-n}(1-\varepsilon_b)\left(\Gamma(\tau)-\Gamma^{*}\right)\vert  e_n^TPy(\tau)\vert   
    \]
Since every symmetric matrix has a unique Cholesky decomposition \cite{horn1990}, i.e., $P=RR^T>0$, with $R$ a lower triangular matrix with real and positive diagonal entries, hence
\begin{align*}
    \vert y^T(\tau)Pe_n \vert = \vert y^T(\tau)RR^Te_n \vert=\vert \bar{y}^T(\tau) {B} \vert
\end{align*} where the vectors $\bar{y}(\tau)=R^Ty(\tau)$ and $B=R^Te_n$ satisfy $\vert \bar{y}(\tau)\vert=y^T(\tau)Py(\tau)^{1/2}$ and $\vert B \vert=(e_n^T P e_n)^{1/2}$, respectively. By using the Cauchy-Schwarz inequality, the term $\vert  e_n^TPy(\tau)\vert$ accepts the following upper bound
\begin{align*}
   \vert  e_n^TPy(\tau)\vert \leq \vert \bar{y}(\tau) \vert \vert B \vert \leq \vert B \vert V^{1/2}(\tau) 
\end{align*} Hence, using the above expression and Rayleigh-Ritz inequality, 
\begin{align*}
     V_1^{'}(\tau)&\leq -\bar{\alpha}\Vert y(\tau) \Vert^2 + 2\kappa(\tau)^{-n}(1-\varepsilon_b)\bar{\Gamma} \vert  e_n^TPy(\tau)\vert\\
     &\leq - c_0V_1(\tau) +2c_1\exp{(-c_2\tau)} V_1^{1/2}(\tau) 
\end{align*} where $c_0=\frac{\bar{\alpha}}{\lambda_{max}(P)}$,  $c_1=\alpha^n T^n(1-\varepsilon_b) \bar{\Gamma} \vert  B\vert$, and $c_2=n\alpha$. The existence of $\bar{\Gamma}>0$ follows from $\Gamma(\tau)$ being bounded, see the arguments in proof of Lemma \ref{lem:pnf}. 
Following the same arguments used in the proof of Claim 11 in \cite{cruz2023uniform}  it can be concluded that there exist $c_5>0$ such that $\Vert y(\tau)\Vert \leq c_5$. Since $\kappa({\tau})^{n}\leq \kappa(\bar{\tau})^{n}<\infty$ is bounded, then the product $\underline{b}^{-1}\kappa(\tau)^{n}\vert u_0(\tau) \vert \leq \underline{b}^{-1}\kappa(\bar{\tau})^{n}\Vert e_n\Vert \Vert P \Vert \Vert y(\tau) \Vert$ is also bounded. Since the time scalings are invertible, for each $\tau \in [0,\: \bar{\tau}]$ the same bound for $y(\tau)$ holds for $x(t)$ in the interval $t\in [0,\bar{t}]$. Thus, one can find $c_\kappa>0$
such that $\underline{b}^{-1}\kappa(t)^n\vert u_0(t) \vert \leq \underline{b}^{-1}c_\kappa$ on $t\in [0,\bar{t}]$.

\noindent \textbf{Step 2.}
The boundedness of the second summand follows from the Proof of Lemma \ref{lem:pnf}, it is shown that there exists a constant $c_6>0$ such that $\vert \Gamma(t)-\Gamma^*\vert <c_6$, thus the following bound holds $\vert \Gamma(t)\vert \leq \vert \Gamma(t) -\Gamma^*\vert + \vert \Gamma^*\vert <c_6 + \Gamma^*$. The second summand can be bounded by $\underline{b}^{-1}\Lambda(t,x(t))\leq \underline{b}^{-1}(c_6 + \Gamma^*)<\infty $ on $t\in [0,\bar{t}]$.

From {Step 1} and {Step 2}, one can conclude that the control law accepts the upper bound $\vert u(t) \vert\leq \underline{b}^{-1}(c_\kappa+c_6 + \Gamma^*)<\infty$ on $t\in [0,\bar{t}]$.
\subsection{Proof of Lemma \ref{lem:barrier}}\label{app:barrier}
Consider the Lyapunov function candidate $V$ along the trajectories of the system
 \begin{equation}
 \begin{aligned}
 \dot{x} &=  J_n x + e_nf(t) \\ &- e_n  (1+\delta_b(t))(\tilde{\gamma} e_n^TPx+ k(V)\text{sign}( \tilde{\gamma}e_n^TPx))   \,,
 \end{aligned}
 \end{equation}
  \normalsize
where $\tilde{\gamma} = \tfrac{1}{2}\left( \tfrac{1}{\gamma}+1 \right)$, yields
 \begin{multline}
 \dot{V} = 2x^TP\left[J_n x + f(t) \right]\\
 -2x^TP e_n  \left[ (1+\delta_b(t))(\tilde{\gamma} e_n^TPx + k(V)\text{sign}(\tilde{\gamma} e_n^TPx))\right]    
 \end{multline}
Similarly that in the proof of Proposition \ref{lemma:lineal}, one can prove that the time derivative of $V$ along the trajectories of $\dot{x} =  J_n x - e_n\tilde{\gamma} (1+\delta_b(t)) e_n^TPx $ satisfies $\dot{V} \leq -\lambda_{min}(Q)\Vert x(t)\Vert^2$ with $P$ solution of \eqref{eq:riky}. Defining $ {\Phi} = \frac{ M }{ 1 - \varepsilon_b }$ and the barrier function $k(V)= \frac{V}{\epsilon - V}$, the derivative of $V$ has the form
 \begin{equation}\label{eq:dvbf}
 \begin{split}
 \dot{V} \leq&  -\lambda_{min}(Q)\Vert x(t) \Vert^2 \\ &- 2(1 - \varepsilon_b) \left\vert x^TPe_n \right\vert \left[  \frac{ V}{\epsilon - V} - {\Phi} \right]\,.
 \end{split}
 \end{equation}
Since $\epsilon-V >0$ as $V$ tends to $\epsilon$, then $\left\vert \frac{ V}{\epsilon - V} - {\Phi} \right\vert < R_V:= 1 + \Phi$, then it is implied by \eqref{eq:dvbf} that $\dot{V} \leq -\beta_1 V(x) + \beta_2V^{\frac{1}{2}} $ with $\beta_1 = \lambda_{min}(Q)/\lambda_{max}(P)$ and $ \beta_2 = 2\left[(1-\varepsilon_b)\lambda_{max}(P)R_V\right]/\sqrt{\lambda_{min}(P)}$, which implies that from the first time $t_0$, the trajectories of the system does not escape of the set $ \mathcal{S}_{\frac{\epsilon}{2} }$ in a finite time. Let us define the variable
 \begin{equation}\label{eq:sigma_1}
 \sigma_1 :=  \frac{\Phi }{ 1 + \Phi}\epsilon
 \end{equation}
 from we will prove that $\epsilon > V(x) > \sigma_1$.  From the definition of function $k(V)$, it follows that there exist $k(V) > k(\sigma) = \Phi$, and it is implied that 
 \begin{equation}
 \dot{V} \leq -\Vert x(t)\Vert^2 - (1-\varepsilon_b)\xi \left\vert2 x(t)^TPe_n \right\vert \leq -\Vert x(t)\Vert^2
 \end{equation}
 where $\xi = k(V) - \Phi > 0$. Then, the trajectories of the closed-loop system converge to the level set $V\leq \sigma_1$. 
 This means that for all $t \geq t_0$ the inequality $V \leq \sigma_1$ is satisfied.  By construction of \eqref{eq:sigma_1}, it follows that $V < \epsilon$ for all $t \geq t_0$, this completes the proof.

\end{document}